\documentstyle[12pt]{article}

\def\bendequ#1#2{\begin{equation}\label{#1}#2\end{equation}}

\def\RE{{\rm Re}}
\def\V#1{{\bf #1}}

\def\dj{d\kern-0.8 ex\vrule 
height 1.32 ex depth -1.24 ex width 0.7ex\kern 0.15 ex}
\def\Dj{D\kern-1.65 ex\vrule 
height 0.95 ex depth -0.87 ex width 0.8ex\kern 0.78 ex}
\def\M {m}
\def\jen#1#2{\frac{{\rm d}\,#1}{{\rm d}\,#2}}
\def\dva#1#2{\frac{{\rm d}^2\,#1}{{\rm d}\,{#2}^2}}
\def\d={=}
\def\RE{\Re} \def\IM{\Im} \def\mm{n}

\def\quarter{{\footnotesize\frac{1}{4}}}
\def\dint{{\rm d\,}}
\def\q0{q_0}

\def\sign{{\rm sign}\,}
\def\energ{E}
\def\textind#1{\indent\llap{\hbox to \parindent{#1\hfill}}\ignorespaces}

\begin{document}

\begin{center}

{\Large\bf Quasiparticle Excitation Spectrum of an Isolated Vortex in a High--Temperature Superconductor}
\vskip \baselineskip
Irena Kne\v zevi\'c and Zoran Radovi\'c\\
{\it Department of Physics, University of Belgrade,\\
P. O. Box 368, 11001 Belgrade, Yugoslavia}
\end{center}

\vskip 1.5\baselineskip
%\section*{ABSTRACT}
\centerline{\large Abstract}

\vskip 0.5\baselineskip
The quasiparticle energy spectrum of an isolated vortex in 
a clean layered $d$-wave superconductor is calculated. The Bogoliubov--de Gennes equations are solved perturbatively, within the model of step-variation of the gap function, adjusting the Caroli, de Gennes and Matricon approach for low-lying excitations in cuprates. A large peak in the density of states in a "pancake" vortex is found, 
as a consequence of two-dimensionality and strong coupling.
\vskip 1.5\baselineskip
\vskip .8\baselineskip

\clearpage
%\noindent{\bf 1. Introduction}
\section{Introduction}

Enhancing interest has been focused on quasiparticle spectra properties around vortices. Density of localized states near the Fermi surface in the vortex core in conventional superconductors ($s$--wave pairing, weak coupling, isotropic) is equivalent to the electron density of states of a normal-metal area with radius of the order of the coherence length $\xi\sim 10\,{\rm nm}$.[1-3] High-temperature superconductors (HTS)
are layered, having a cylindrical Fermi surface, 
the superconductivity is of $d$-wave and strong coupling type, the coherence length being much smaller, $\xi\sim 1{\rm nm}$. This implies a crucial difference in the quasiparticle spectra in the vortex core between conventional superconductors and HTS. 

Numerous computations of the quasiparticle spectra around two-dimensional (2D) vortices have been performed recently.[4-10] The common feature of layered  superconductors is the appearance of a large peak in the quasiparticle density of states (DOS) in the vortex core, in agreement with the scanning tunneling microscopy (STM) observations on ${\rm Nb Se_2}$\cite{Hess} and YBCO.\cite{Magio} However, in BSCCO the gap structure is found, without any signature of the quasiparticle bound states in the vortex core.\cite{Renner}

In this paper, the Bogoliubov--de Gennes equations for the vortex core are solved analytically, using the perturbation theory within the model of step-variation of the gap function. Since the bound states spectrum is not quasicontinuous for 2D  and strong coupling superconductivity, this approach modifies the Caroli, de Gennes and Matricon results for low-lying states. Our results may confirm that DOS has either one large maximum or gap structure, depending on the value of $k_F\xi$ in HTS.

\section{The Model Solution}
%\noindent{\bf 2.Bogoliubov--de Gennes equations }

For an isolated "pancake"  vortex in HTS in a low magnetic field applied along the $c$-axis, 
using the cylindrical coordinates $(r,\varphi,z)$, $z\parallel c$, we take the gap function
\bendequ{E8}
{\Delta=\Delta(r,\theta)e\sp{-\imath\varphi},}
due to the flux quantization, neglecting the vector potential contribution, $\V A\approx 0$. 

Furthermore, neglecting the self-consistency condition, we assume the spatial variation of the gap function
\bendequ{B-dG2}
{\Delta=\Delta(r,\theta)=\cases{0,& $r<r_c(\theta)$\cr \Delta (\theta),& $r>r_c(\theta)$},}
where 
\bendequ{dspar}
{\Delta(\theta)=\Delta_0\cos (2\theta)} 
for $d$-wave pairing, $\theta$ being the polar angle in $\V k$-space. The vortex core radius $r_c(\theta)$ is of the order of   the BCS coherence length $\xi_0(\theta)=\hbar v_F/\pi|\Delta(\theta)|$.

Taking for the Bogoliubov amplitudes $u$ and $v$
\bendequ{B-dG21}
{\pmatrix{u(\V r,\theta) \cr v(\V r,\theta)}=
\pmatrix{f_+(r,\theta)e^{\imath(\nu-1/2)\varphi}\cr 
f_-(r,\theta)e^{\imath(\nu+1/2)\varphi}},}
where $2\nu$ is an odd integer, we write the Bogoliubov--de Gennes equations in the form 
\bendequ{B-dG23a}
{\left({\cal H}_0+{\cal H}_1\right)f=\energ f,}
\bendequ{B-dG23}
{{\cal H}_0=-L\hat\sigma_z+\Delta (r,\theta) \hat \sigma_x,\qquad {\cal H}_1=-\frac{\nu\hbar^2}{2mr^2},\qquad f=\pmatrix{f_+\cr f_-},}
\bendequ{B-dG23b}
{L =\frac{\hbar^2}{2m}
\left(\dva{}{r}+\frac{1}{r}\jen{}{r}-\frac{\nu^2+\quarter}{r^2}+k_F^2\right),}
where $\hat \sigma_i$ are the Pauli matrices.

We solve Eqs.(\ref{B-dG23a}) perturbatively. First,\cite{mi1} the eigenproblem ${\cal H}_0f=\energ_0 f$, i.e.
\bendequ{B-dG25}
{\pmatrix{L +\energ_0 & -\Delta(r,\theta)\cr  
\Delta(r,\theta) & 
L -\energ_0 }\pmatrix{f_+ \cr f_-}=0. }
Second, the energy shift according to the first-order 
perturbation theory.

We restrict our consideration to localized states, with energies below the bulk gap, 
$|\energ_0|<|\Delta (\theta)|$. The solutions of Eq. (\ref{B-dG25}) are real-valued, 
finite for $r=0$, and zero for $r\to \infty$, 
\bendequ{B-dG10}
{f_\pm=A_\pm J_{\mm} \left( q_{\pm} r \right),}
for $r<r_c(\theta)$, and for $r>r_c(\theta)$ 
\bendequ{B-dG18}
{f_+=BH_\mm^{(1)}(kr)+B^*H_\mm^{(2)}(k^*r),}
\bendequ{B-dG19}
{f_-=-\imath\frac{\sqrt{\Delta^2(\theta)-\energ_0^2}}{\Delta(\theta)}
\left(BH_\mm^{(1)}(kr)-
B^*H_\mm^{(2)}(k^*r)\right)+
\frac{\energ_0}{\Delta(\theta)}\left(BH_\mm^{(1)}(kr)+
B^*H_\mm^{(2)}(k^*r)\right).}
Here, $J$ and $H$ are the Bessel and the Henkel function, 
respectively, $A_\pm$ are real and $B$ is a complex constant, 
$q_{\pm}=\sqrt{k_F^2\pm 2\M\energ_0/\hbar^2}$, and 
$k=\sqrt{k_F^2+\imath\sqrt{2\M(\Delta^2(\theta)-\energ_0^2)/\hbar^2}}$.

Continuity of $f_+$, $f_-$ and their first derivatives at $r=r_c(\theta)$ 
yields the system of algebraic equations for $A_\pm$, $\RE B$ and $\IM B$, 
which  has a nontrivial solution if
\bendequ{B-dG19a}
{\small \left|\matrix
{J_\mm(+)&0 &-\RE H_\mm^{(1)}&\Im H_\mm^{(1)}\cr
0& J_\mm(-)&-\frac{{\cal E}_0}{\Delta(\theta)}\IM H_\mm^{(1)}-
\frac{E_0}{\Delta(\theta)}\RE H_\mm^{(1)}&
-\frac{{\cal E}_0}{\Delta(\theta)}\RE H_\mm^{(1)}+
\frac{E_0}{\Delta(\theta)}\IM H_\mm^{(1)}\cr
q_+{J_\mm}'(+)&0& 
-\RE (k{H_\mm^{(1)}}')&\IM (k{H_\mm^{(1)}}')\cr
0&q_-{J_\mm}'(-)& 
-\frac{{\cal E}_0}{\Delta(\theta)}\IM (k{H_\mm^{(1)}}')-\frac{E_0}{\Delta(\theta)}
\RE (k{H_\mm^{(1)}}') &
-\frac{{\cal E}_0}{\Delta(\theta)}\RE (k{H_\mm^{(1)}}')+\frac{E_0}{\Delta(\theta)}\IM (k{H_\mm^{(1)}}')}\right|=0.}
Here ${\cal E}_0=\sqrt{\Delta^2 (\theta) -E_0^2}$, $J_\mm(\pm)=J_\mm(q_{\pm}r_c)$, $H_\mm^{(1)}=H_\mm^{(1)}(kr_c)$,  
$J'(x)=\dint J(x)/\dint x$, $H'(x)=\dint H(x)/\dint x$. Energies $E_0$ are calculated numerically from Eq. 
(\ref{B-dG19a}). In the first order of the perturbation theory, we obtain the energy of the localized states $\energ =\energ_0+\delta \energ$, where the energy shift is
\bendequ{B-dG30}
{\delta \energ =
-\nu\frac{\hbar^2}{2\M}\frac{\int_0^\infty  \left(f_+^2+f_-^2\right)r^{-1}\dint r}
{\int_0^\infty \left(f_+^2+f_-^2\right)r\dint r}.}
Density of localized states
\bendequ{dos}
{N(E)=\frac{1}{2\pi}\int_0^{2\pi}\dint\theta\sum_\nu
\delta\left(E-E_\nu (\theta)\right),}
is calculated approximately for $E>0$,\cite{mi1} taking  
\bendequ{approx1}
{{E_\nu}(\theta)\approx {E_{0\nu}(0)|\cos(2\theta)|+\delta E_\nu(0)\cos^2(2\theta)}.}

\section{Discussion}
%\noindent{\bf 3. Results and Discussion}

Results are illustrated in Figs.1-3. For a $d$-wave "pancake" vortex, only the lowest localized state ($\nu =1/2$) is significantly separated from the gap edge for $k_F\xi_0(0)\sim 1$ (Fig. 1). The energy shift, Eq. (\ref{B-dG30}), of higher states ($\nu>5/2$) is practically negligable. Density of the lowest localized states ($\nu\leq 5/2$), calculated within this model, is presented in Fig. 2. Only the lowest peak is considerably separated from the gap edge, because of the small $k_F\xi_0(0)$. One large peak is distinct from the gap edge for $0.7<k_F\xi_0(0)<1.2$, which may be the case for YBCO, and none (gap structure) for $k_F\xi_0(0)<0.7$, as for BSCCO, which is in consistency with larger gap for the latter (Fig. 3). This can be the simplest explanation of the experimental observations.\cite{Magio,Renner}

In conclusion, an approximate analytical method is developed in the same manner as the Caroli, de Gennes and Matricon approach, adjusted for the case of $d$-wave pairing and small radius pancake vortices, where the quasiparticle spectrum is not quasicontinuous. Two--dimensional and strong coupling nature of superconductivity in cuprates makes the density of localized quasiparticle states in the vortex core completely unlike the normal-metal DOS. 

\clearpage

\section{Figure Captions}

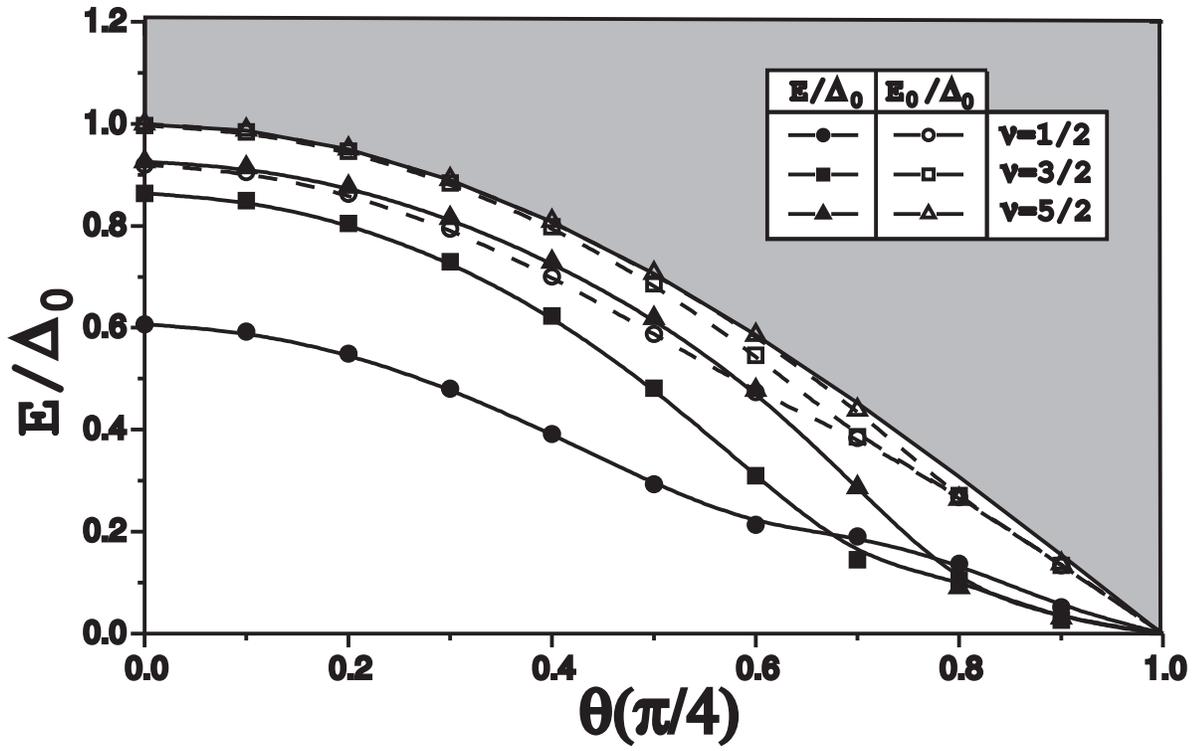
Fig. 1. Quasiparticle energy $E/\Delta_0$ vs angle $\theta$, in the vortex core of radius $r_c=\xi_0(\theta)$, for $k_F\xi_0(0)=1.5$. Dashed curves: $E_0/\Delta_0$. Shaded region: unbound states.

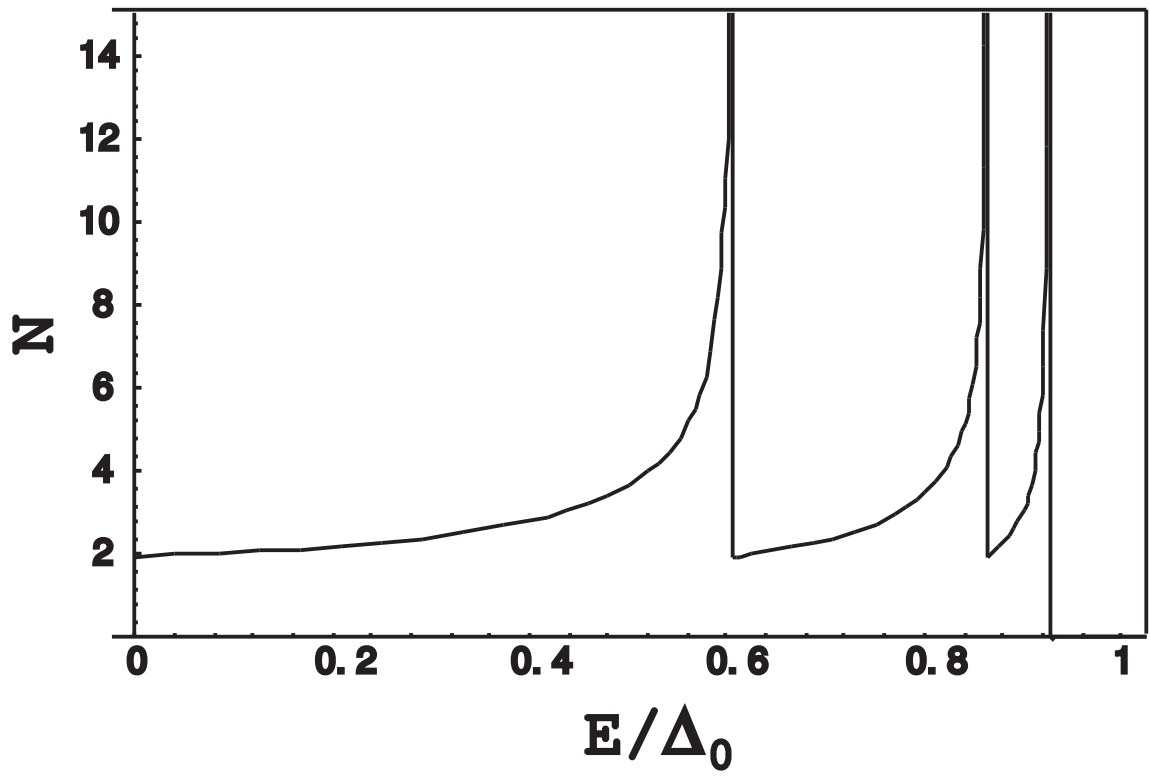
\noindent Fig. 2. Density of localized states in the vortex core of radius $r_c=\xi_0(\theta)$, for $k_F\xi_0(0)=1.5$. 

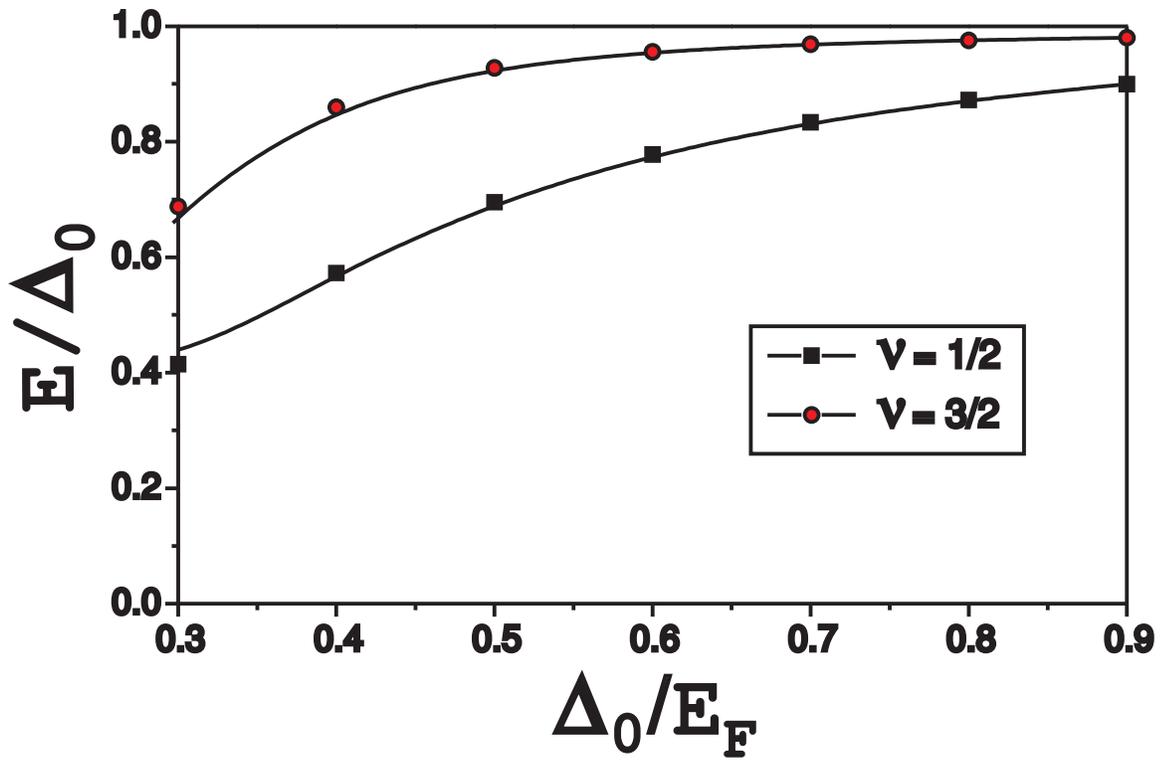
\noindent Fig. 3. Quasiparticle energy of two lowest bound states for $\theta=0$ as a function of $\Delta_0/E_F = 2/\pi k_F\xi_0(0)$.

\end{document}